\documentclass{article}
\usepackage[utf8x]{inputenc}
\usepackage{amsmath} \usepackage{amsthm} \usepackage{amsfonts} \usepackage{amssymb}
\usepackage{array} 
\usepackage{booktabs}
\usepackage{enumitem}
\usepackage{indentfirst}
\usepackage{latexsym}
\usepackage{makeidx}
\usepackage{graphicx}
\usepackage{booktabs}
\usepackage[font=small,format=hang,labelfont=sf]{caption}
\usepackage{subfig}
\theoremstyle{plain} 
\usepackage[rightcaption]{sidecap}

\newtheorem{thm}{Theorem}[section] 
\theoremstyle{definition} 
\newtheorem{lem}[thm]{Lemma} 

\begin{document}
\title {Supplementary material for \\``Dynamical correlations in the escape strategy of Influenza A virus''} 
\author{Lorenzo Taggi$^{1,2}$, Francesca Colaiori$^{1,3}$, Vittorio
  Loreto$^{1,4}$ and Francesca Tria$^4$}

\maketitle
\centerline{$^1$Sapienza University of Rome, Physics Department, P.le
  A. Moro 5, 00185 Rome, Italy}

\centerline{$^2$ Max Planck Institute for Mathematics in the Sciences,
  Inselstr. 22, 04103 Leipzig, Germany}

\centerline{$^3$ CNR-ISC, P.le A. Moro 5, 00185 Rome, Italy}

\centerline{$^4$ ISI Foundation, Via Alassio 11/c, 10126, Torino,
  Italy}

\section{Properties of the Epistatic Immunity Set}
\subsection{Size}
As discussed in the main text, in our model the immunity set
$I_n(\vec{v})$ is defined as the number of strings that do not contain
two adjacent mutations with respect to a given string $\vec{v}$ (we
consider ``adjacent'' also the first and the last bit of every
string),
\begin{equation}
\label{math:EIset}
I_n(\vec{v})= \{\  \vec{z}\in H_n :\  z_{i}\neq v_{i} \Rightarrow \ z_{|i+1|_n}= v_{|i+1|_n}\  \forall i  \ \}.
\end{equation}

\noindent Since cross-immunity, as defined in (\ref{math:EIset}), is
only sensible to differences between strains, $I_n(\vec{v})$ is
invariant under translation on the hypercube with periodic boundary
conditions. In order to see this, we can define a sort of
translational operator acting on the space of sequences,
\begin{equation}
\label{math:translationoperator}
 \mathcal{X}_{\vec{c}}\, (\, \cdot \,) \, =
\, \vec{c} \ \ \veebar \ \cdot~~,
\end{equation} 

\noindent where $\vec{c}$ is a generic strain and ``$\,\veebar\,$'' is
the \textit{XOR} operator. Both the EI-sets and the hamming distance
are invariant to it:

\begin{gather} 
  \label{math:invariance1}
  \vec{v}_{2} \in I_n(\vec{v}_1) \Leftrightarrow \mathcal{X}_{\vec{c}} (\vec{v}_{2}) \in I_n(\,  \mathcal{X}_{\vec{c}} (\vec{v}_1)\, ),\\
\label{math:invariance2}
d ( \vec{v}_1, \vec{v}_2 ) = d(\, \mathcal{X}_{\vec{c}}( \vec{v}_1)\,
,\, \mathcal{X}_{\vec{c}} (\vec{v}_2) \,) \,\,\, \forall \vec{c}.
\end{gather}

\noindent Therefore $I_n(\vec{v})=\mathcal{X}_{\vec{v}}
(I_n(\vec{0}))$. Then, in order to characterize the static properties
of $I_n(\vec{v})$, we can take the null vector $\vec{0}$ as generating
strain for the immunity set without loss of generality. Let us now
compute the cardinality of the immunity set. In order to do this, we
first compute the cardinality of the EI set ``without boundary
conditions'' (EISNB), defined as:

\begin{equation}
  \label{math:EISNB}
  I^{NB}_n(\vec{v})= \{\  \vec{z}\in H_n :\  z_{i}\neq v_{i} \Rightarrow
  \ z_{i+1}= v_{i+1}\ \, \, \forall i \in {0, 1, ... , n-1}  \ \},
\end{equation}

\noindent and then we show that the cardinality of (\ref{math:EIset})
is a linear combination of cardinalities of EISNB sets.

\begin{lem}
\label{lemma_1.1}
  Let us call $B(n)$ the cardinality of the EISNB. This number follows
  the recursive law,
$$
B(n)=B(n-1)+ B(n-2),
$$
with initial conditions $B(0)=1$ and $B(1)=2$.			 
\end{lem}

\proof 

\noindent
\begin{itemize}  
\item {$I^{NB}_{1}(\vec{0}) = \{ (0), (1)\} \Rightarrow B(1)=2$}. 
\item{ $I^{NB}_{2}(\vec{0}) = \{ (0, 0), (1, 0), (0, 1) \} \Rightarrow B(2)=3$}.  
\item{ Let's consider the set $ I^{NB}_{n}(\vec{0})$ for a generic
    dimension $n$. This set is equivalent to the union of the two
    disjoint sets $C_{n}$ and $D_{n}$, which are respectively defined
    as the set of all strings belonging to $I^{NB}_{n}(\vec{0})$ with
    the first bit equal to $1$ and as the set of all strings belonging
    to $I^{NB}_{n}(\vec{0})$ with the first bit equal to $0$:

    \begin{equation} \label{math:eq1} C_{n}= \{ \vec{v} \in H_n \, \,
      \, s.t. \, \, \, \vec{v} = (1, 0, \vec{c}), \, \, \, \vec{c} \in
      I_{n-2} (\vec{0}) \},
    \end{equation}

    \begin{equation}
      \label{math:eq2} D_{n}= \{ \vec{v} \in H_n \, \, \,
      s.t. \, \, \, \vec{v} = (0, \vec{c}) , \, \, \, \vec{c} \in I_{n-1}
      (\vec{0}) \}.
    \end{equation} 

    \noindent In equation (\ref{math:eq1}) we have considered that, by
    definition of EISNB, if the first bit is mutated, the second one
    cannot be mutated. It's easy to see that the cardinality of the
    first set is equal to $B(n-2)$ and that the cardinality of the
    second one is equal to $B(n-1)$. Being the two sets disjoint,
    $B(n)=B(n-1)+ B(n-2)$.}
\end{itemize}
\endproof

\noindent Thus, the cardinality of the EISNB follows the well known
Fibonacci rule. We will use this result to determine the cardinality
of the Epistatic Immunity Set.

\begin{lem}
Let us call $S(n)$ the cardinality of the Epistatic Immunity Set. This number follows the following recursive law,
$$
S(n)=S(n-1) +S(n-2),
$$
with initial conditions $S(2)=3$ and  $S(3)=4$.
\end{lem}
\proof
\noindent
As for the previous lemma, the proof is by induction.
\begin{itemize}
\item  $I_{2}(\vec{0}) = \{ (0, 0), (1, 0),\, (0\, 1) \} \Rightarrow S(2)=3$; 
\item  $I_{3}(\vec{0}) = \{ (0, 0, 0), (1, 0, 0),\, (0, 1, 0),\, (0, 0, 1) \} \Rightarrow S(2)=4$;
\item  We observe that the following relation between the EI set and
  the EISNB holds: 

\begin{equation}  
I_{n}(\vec{0}) = I^{NB}_{n}(\vec{0})\ \ \setminus\ \ \{ \vec{z} \in
  I^{NB}_{n}(\vec{0})\ :\ \vec{z} = (1, 0, \vec{c}, 0, 1), \ \ \vec{c}
  \in I_{n-4}^{s}(\vec{0}) \}.
\end{equation}

\noindent Then, 

\begin{equation} 
  \label{math:eq3} S(n) = B(n) - B(n-4). 
\end{equation}

\noindent Applying Lemma~\ref{lemma_1.1} to equation (\ref{math:eq3})
one gets:

\begin{equation} \label{math:eq4} B(n) - B(n-4) = B(n-1) + B(n-2) -
  B(n-5) - B(n-6).
\end{equation} 

\noindent Then, collecting the terms properly and using again equation
(\ref{math:eq3}), we finally find the desired relation: $$S(n) =
S(n-1) + S(n-2).$$
\end{itemize}
\endproof

\noindent The sequence $S(n)=S(n-1) + S(n-2)$, under the initial
condition specified by the previous lemma, is called \textit{Lucas
  sequence}. As for the Fibonacci sequence, the fraction of two
consecutive numbers of the Lucas sequence converges asymptotically to
the value $\Phi = \frac{1 + \sqrt{5}}{2}$, which is well known as the
{\itshape Golden Ratio}. In particular it is easy to show that $S(n)=
{\Phi}^{n} + (1-{\Phi})^{n} \thicksim\ \Phi^{n}$.

\subsection{Density}

\begin{figure}[!htb]
\begin{center}
\includegraphics[scale=0.41]{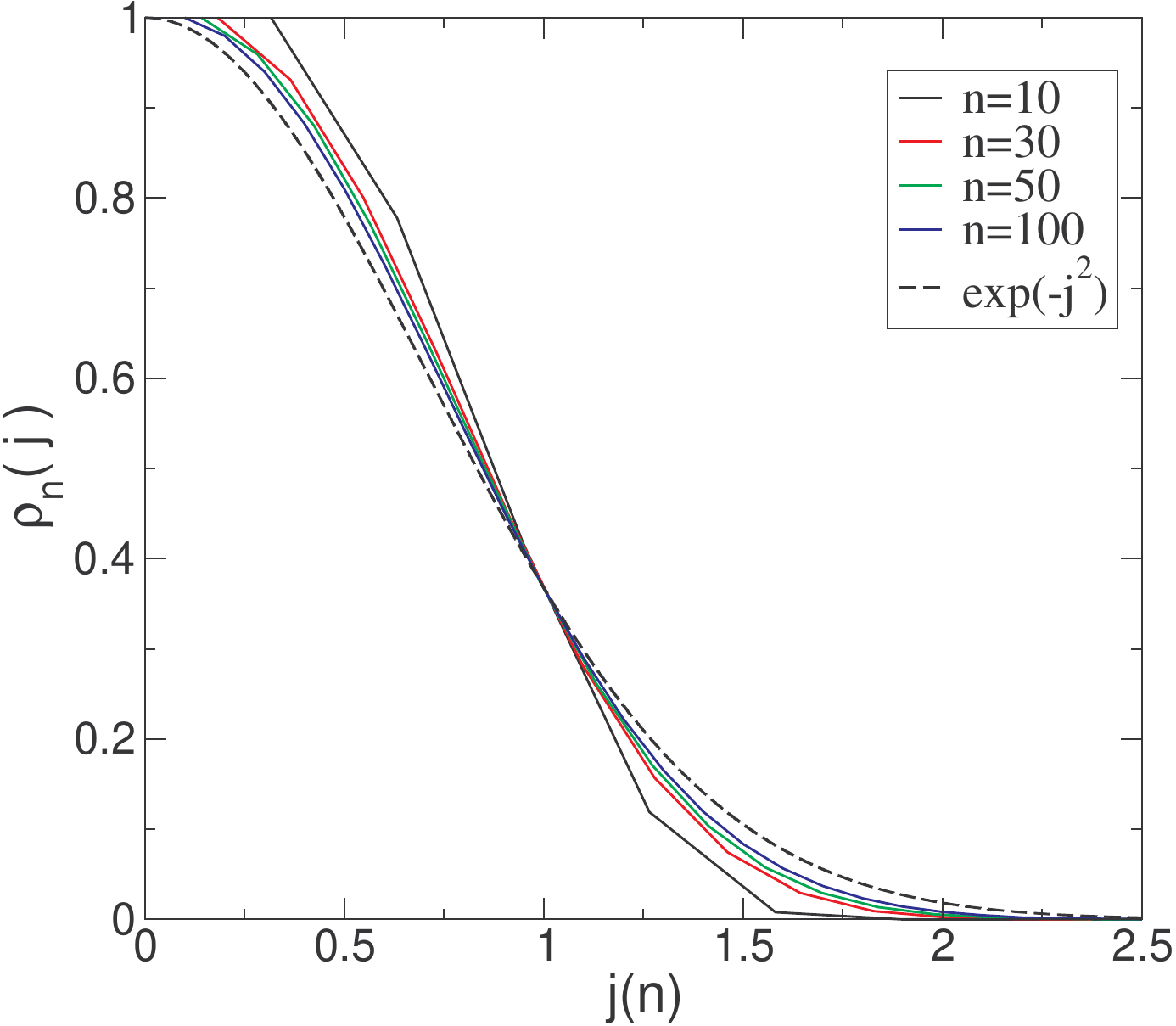}
\caption{Epistatic density function computed numerically directly from
  definition (\ref{math:rho}) for $n=10, 30, 50, 100$. The densities
  are plotted as function of $j(n) = i/\sqrt{n}$, where $i$ is the
  length. As $n$ increases, the epistatic density function converges
  to the well defined function $\rho_{\infty}(j) = exp(-j^{2})$.}
\label{fig:densepi}
\end{center}
\end{figure}

We define the \textit{Epistatic density function} as the ratio between
the number of strings, $L(n,i)$, contained in $I_n(\vec{0})$ and having
hamming distance $i$ from $\vec{0}$ and the number $V(n,i)$ of strings
having hamming distance $i$ from $\vec{0}$:
\begin{equation}
\label{math:rho}
\rho_{n}(i) \, := \frac{L(n,i)}{V(n,i)}.
\end{equation}

\noindent This function gives an idea of how the elements of the
epistatic immunity set are distributed on the hypercube. It is easy to
check that $L(n,i)=\binom{n - i + 1}{i} - \binom{n - i - 1}{i - 2}$. The
first term represents the number of strings not containing two
adjacent ones; the second term represents the number of strings not
containing any pairs of adjacent ones, but with the first and the last
bit both equal to one. Thus the second term takes into account the
effect of the periodic condition in definition (\ref{math:EIset}). The
denominator is simply $V(n,i)= {\binom{n}{i}}$. The density function
(\ref{math:rho}), computed numerically, is represented for different
values of $n$ and plotted as function of $i/\sqrt{n}$ in Fig.
\ref{fig:densepi}. We see that  $\rho_n(i)$ can be approximated as
\begin{equation}
\label{math:densepi4}
\rho_{n}(i) \, \simeq \, \exp({-\frac{i^2 }{ n}}),
\end{equation}
and, substituting $j = i / \sqrt{n}$, we get
\begin{equation}
\label{math:densepi15}
\rho_{n}(j) \simeq  \exp(-j^2).
\end{equation}
\noindent Therefore, the epistatic immunity set covers an area of the
hypercube whose size grows proportionally to $\sqrt{n}$. In this area,
the density of strings satisfies equation (\ref{math:densepi4}).
We now prove analytically validity of approximation (\ref{math:densepi4})
in the range $i \leq \sqrt{n}$. 
\begin{equation}
\label{math:densepi2}
\begin{split}
\rho_{n}(i) &= \frac{(n-i)! \cdot (n-i+1)!}{n! \cdot (n - 2i + 1)!}  [ \, 1 - \frac{i \cdot (i-1)}{(n - i + 1) \cdot (n - i) } \, ] \\
            &= \frac{(n-i)\cdot (n - i - 1) \ldots (n - 2i + 2)}{n \cdot (n-1) \ldots (n - i + 2)}          \  [\, 1 + {O(i/n)}^2\, ] \\
            &= ( 1 - \frac{i}{n}) \cdot (1 - \frac{i}{n-1})\ \ldots\ (1 - \frac{i}{n - (i - 2)} )\, \cdot\, (\ 1 \ +\  {O(i/n)}^2 \ ) \\
            &= { (\  1 - i/n + {O(i/n)}^2 \ )}^{i-1} \cdot (\ 1 \ +\  {O(i/n)}^2 \ ) .
\end{split}
\end{equation}

\noindent Using the expansion:

\begin{equation}
  \label{math:expansion}
  {(1 + x)}^{i-1} = \sum_{l=0}^{i-1}\binom{i-1}{l}x^{l},
\end{equation}

\noindent one gets:

\begin{equation}
\begin{split}
  &=  \left[\sum_{l=0}^{i-1} \binom{i-1}{l} {[ {O(i/n)}^2 - i/n ]}^l \right] \cdot (\, 1 \, + \, {O(i/n)}^2 \, )\\
  &=  \left[\sum_{l=0}^{i-1} \binom{i-1}{l} [ -{(i/n)}^l\,  +\,  {O(i/n)}^{l+1} \, + \, {O(i/n)}^{l+2} \, \ldots {O(i/n)}^{2l}] \right] \cdot (\, 1 \, + \, {O(i/n)}^2 \, )\\
  &= \lbrace 1\ +\ [\, - \frac{i^2}{n} + O(\frac{i}{n}) \, ]\ +\
  [\, \frac{i^4}{2! n^2} + O(\frac{i^3}{n^2}) \, ] \ +\ [ -
  \frac{i^6}{3! n^3} + O(\frac{i^5}{n^3}) \,]\ + \ldots \rbrace \cdot
  (\, 1 \, + \, {O(i/n)}^2 \, ).
\end{split}
\end{equation}

\noindent Then, substituting $j= i / \sqrt{n}$ and considering $j \leq 1$, we get equation (\ref{math:densepi15}):

\begin{equation}
\begin{split}
  \rho_{n}(j) = O(\frac{j}{\sqrt{n}}) + 1 - j^2 + \frac{j^4}{2!} - \frac{j^6}{3!} \ldots  \\
  = \exp(-j^2) + O(\frac{j}{\sqrt{n}}).
\end{split}
\end{equation}

\section{Numerical estimate of m(n)}

\begin{figure}[!htb]
\centering
\includegraphics[width=0.6\columnwidth]{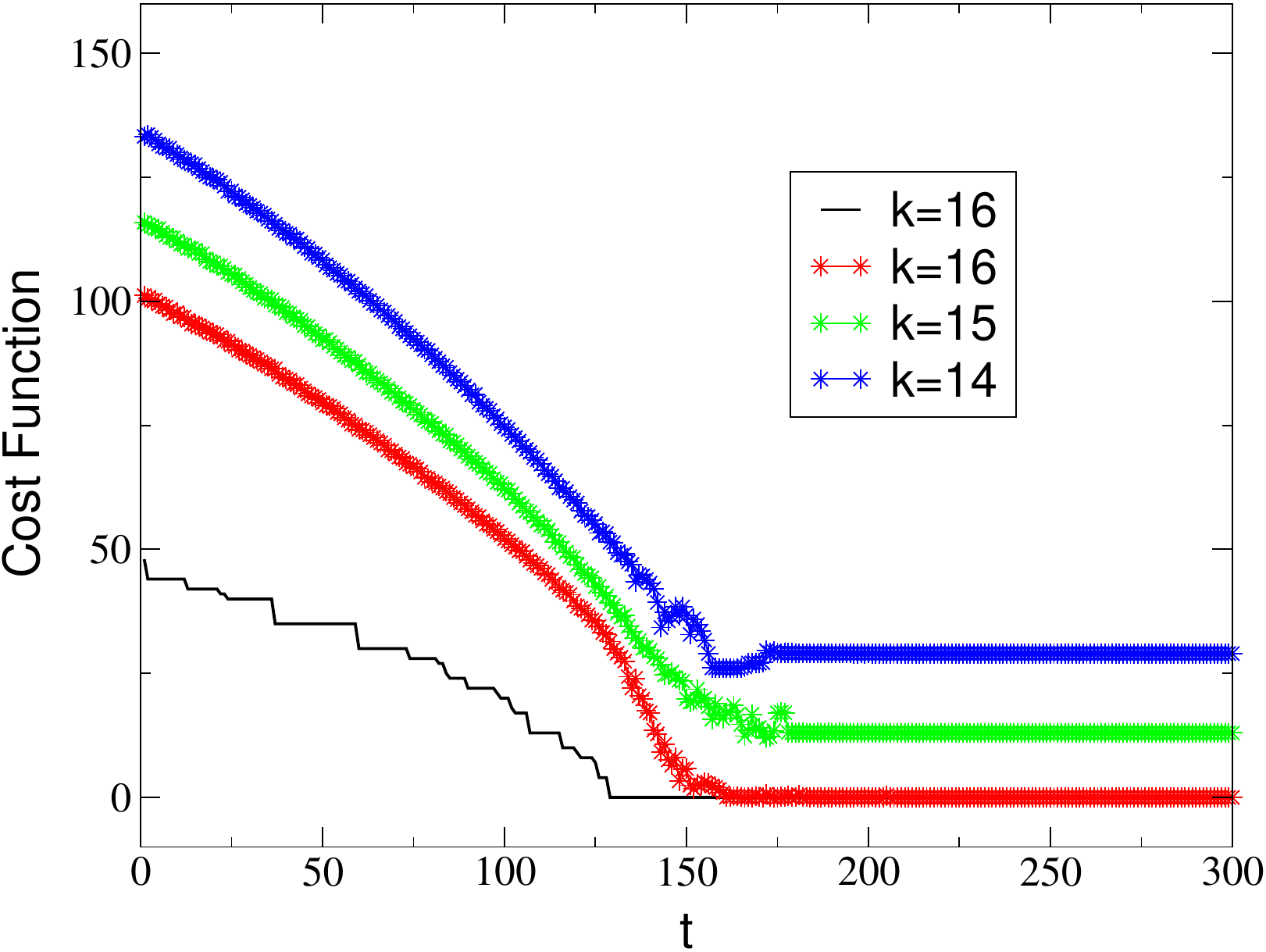} 
\includegraphics[width=0.6\columnwidth]{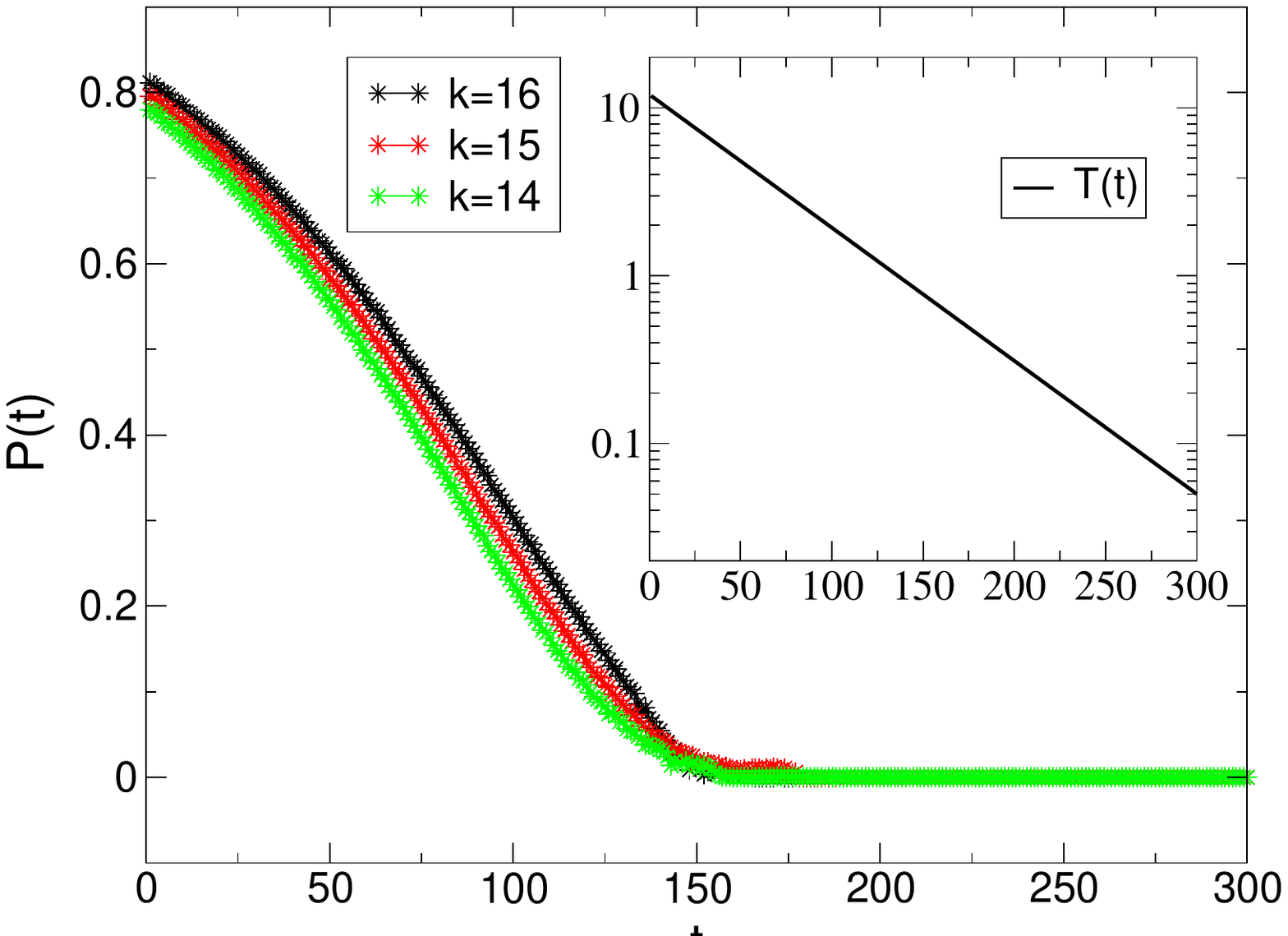}
\caption{\textbf{Top.} Cost function averaged over several
  realizations for the same temperature $T(t)$ and for $k=14$, $15$,
  $16$ and $n=10$ (\textit{dots}); minimum value of the cost function
  obtained until time $t$ for $k=16$ and $n=10$ (\textit{continuous
    line}). \textbf{Bottom.} The acceptance probability $P(t)$, which
  is equal to the fraction of solutions accepted at temperature
  $T(t)$. \textbf{Inset.} The thermal function adopted for these
  simulations, $T(t)=T_0 \cdot \alpha^t$, with $\alpha\simeq 0.982$
  and $T_0=15$.}
\label{fig:sumulatedannealing}
\end{figure}

In the main text we introduced the quantities $m(n)$ and $M(n)$ in
order to investigate how the introduction of a correlated rule for
cross-immunity shapes the EIS. $m(n)$ is defined as the minimum number
of strings needed to cover with their immunity sets the whole
sequences space; $M(n)$ is the maximum number of distinct strings that
can be accommodated in the space of sequences still leaving some strings
out of their EIS. Essentially, in the two cases the set of strings
that realizes the minimum (maximum) will be such to minimize
(maximize) the overlap among immunity sets. In the main text we have
computed $M(n)$ analytically and we have provided analytical and
numerical estimates for $m(n)$. In order to compute the numerical
estimate for $m(n)$ we adopted a Simulated Annealing
approach~\cite{Kirkpatrick}.

\begin{figure}[!htb]
  \begin{center}
    \includegraphics[scale=0.35]{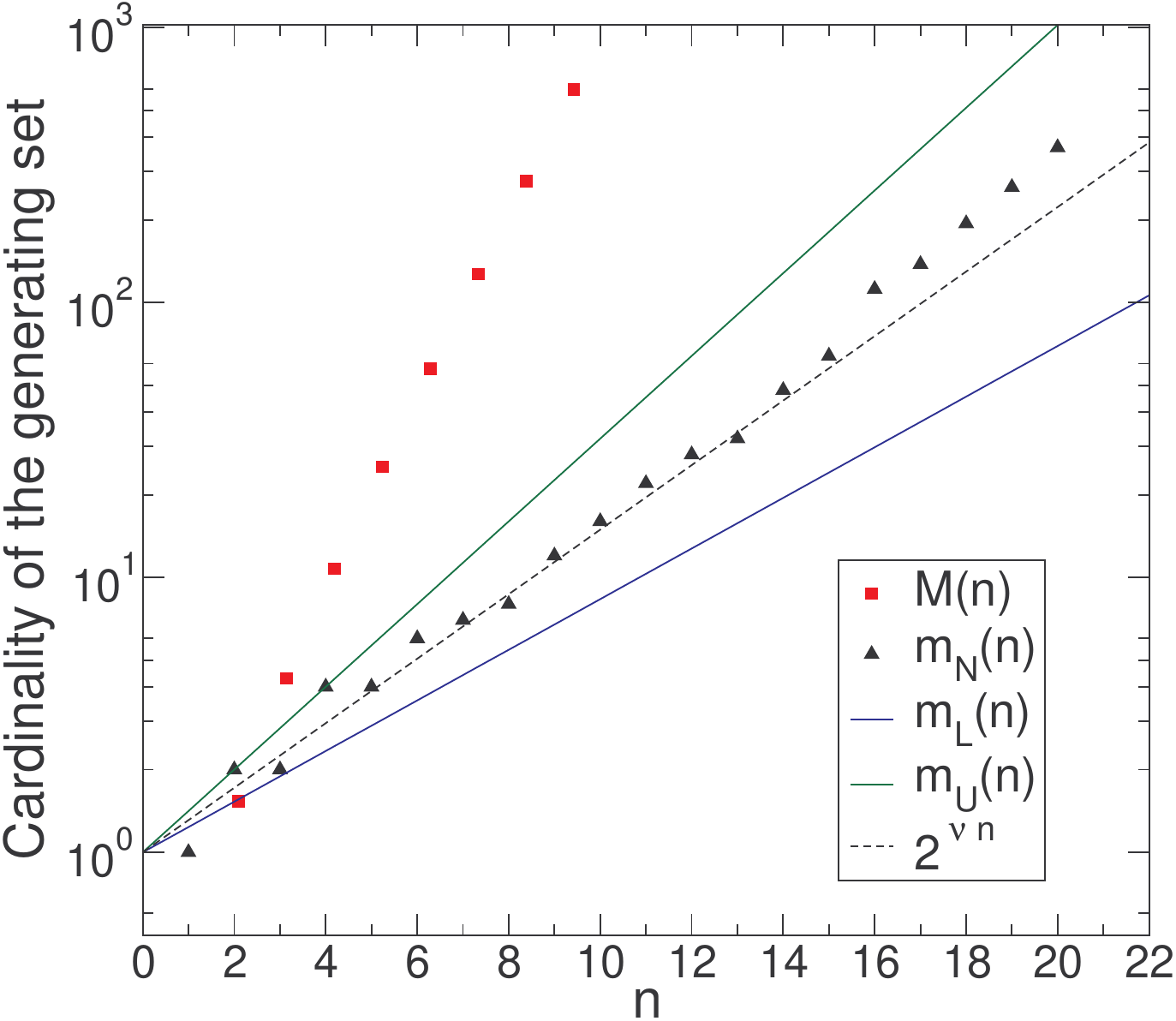}
    \caption{Numerical estimate of $m(n)$, $m_N(n)$, along with its lower and
      upper bound, and the value of $M(n)$ as a function of $n$. The
      dotted line represents the function $2^{\nu n }$, with $\nu =
      0.399 \pm 0.002$, which has been used to fit the first $15$
      values of $m_N(n)$.}
    \label{fig:stime}
  \end{center}
\end{figure}

We first notice that an exhaustive search of the solution would not be
possible because the space of the infection sets is too big: for
instance, the number of possible infection sets with cardinality $16$
in a space of dimension $10$ is $\binom{~2^n}{k} = \binom{~2^{10}}{16}
\sim 10^{35}$. We instead proceeded as follows: we fix a value $k$ for
the number of elements of the infection set $A$ and we search for a
configuration which minimizes the {\em cost function} $E_{n,k} (A_k)
= 2^n -|I_n(A_k)|$ (we denominate $A_k$ any infection set with
cardinality $k$). Of course, $E_{n,k}(A_k) \geq 0$ for all $k$: the
smallest $k$ for which we obtain $E_{n,k} (A_k ) = 0$ for some set
$A_k$ is the numerical estimate for $m(n)$, $m_N(n)$, and the relative set is one of the
possible $A_{min}$. For fixed values of $n$ and $k$ we start from a
random choice of the infection set $A_k$ and we modify a randomly
selected bit of a randomly selected string in $A_k$. Then we compute
the new cost function $E_{n,k} (A^{\prime}_k)$ and

\begin{itemize}
\item if $E_{n,k}(A^{\prime}_k) < E_{n,k}(A_k)$, then $A_k$ is
  replaced by $A^{\prime}_k$ with probability $1$;
\item if $E_{n,k}(A^{\prime}_k) \geq E_{n,k}(A_k)$, then $A_k$ is
  replaced by $A^{\prime}_k$ with probability $P(\Delta E) = \exp( -
  \Delta E / T(t))$.
\end{itemize}

\noindent We iterate this procedure $R(t)$ times for every value of
the temperature $T(t)$. Afterwards, temperature is updated with the
rule $T(t+1) = \alpha \cdot T(t)$, where $0<\alpha<1$. The number of iterations performed
for every temperature value, $R(t)$, is chosen to grow exponentially
with $t$. In fact, recalling the analogy with Statistical
Mechanics~\cite{Kirkpatrick}, lower is the temperature, larger is the
time a body that can exchange heat with a thermal bath needs to reach
the thermal equilibrium. In Fig.~\ref{fig:sumulatedannealing} (top) we
report the behaviour of the cost function, averaged over all the
iterations performed at the same temperature $T(t)$, as function of
$t$. In Fig.~\ref{fig:sumulatedannealing} (bottom) we report the
fraction of accepted solutions at time $t$ as well as the function
$T(t)$ considered. When the average cost function and the minimum cost
function stop decreasing and remain constant, a local minimum is
reached.

\noindent The results obtained for the cardinality of the generating
set are reported in Fig.~\ref{fig:stime} for values of $n$ up to
$n=20$. We also report the upper and lower bounds for $m(n)$. Due to
the high computational complexity of the problem (the number of local
minima of the cost function and the size of the solutions space both
grow very fast with $n$ and $k$), our numerical estimate of $m(n)$
tends to be rougher for higher values of $n$.

\section{Cluster structure of the EIS}
\label{sec:clusters}

\begin{figure}[!htp]
\centering
\includegraphics[width=0.86\columnwidth]{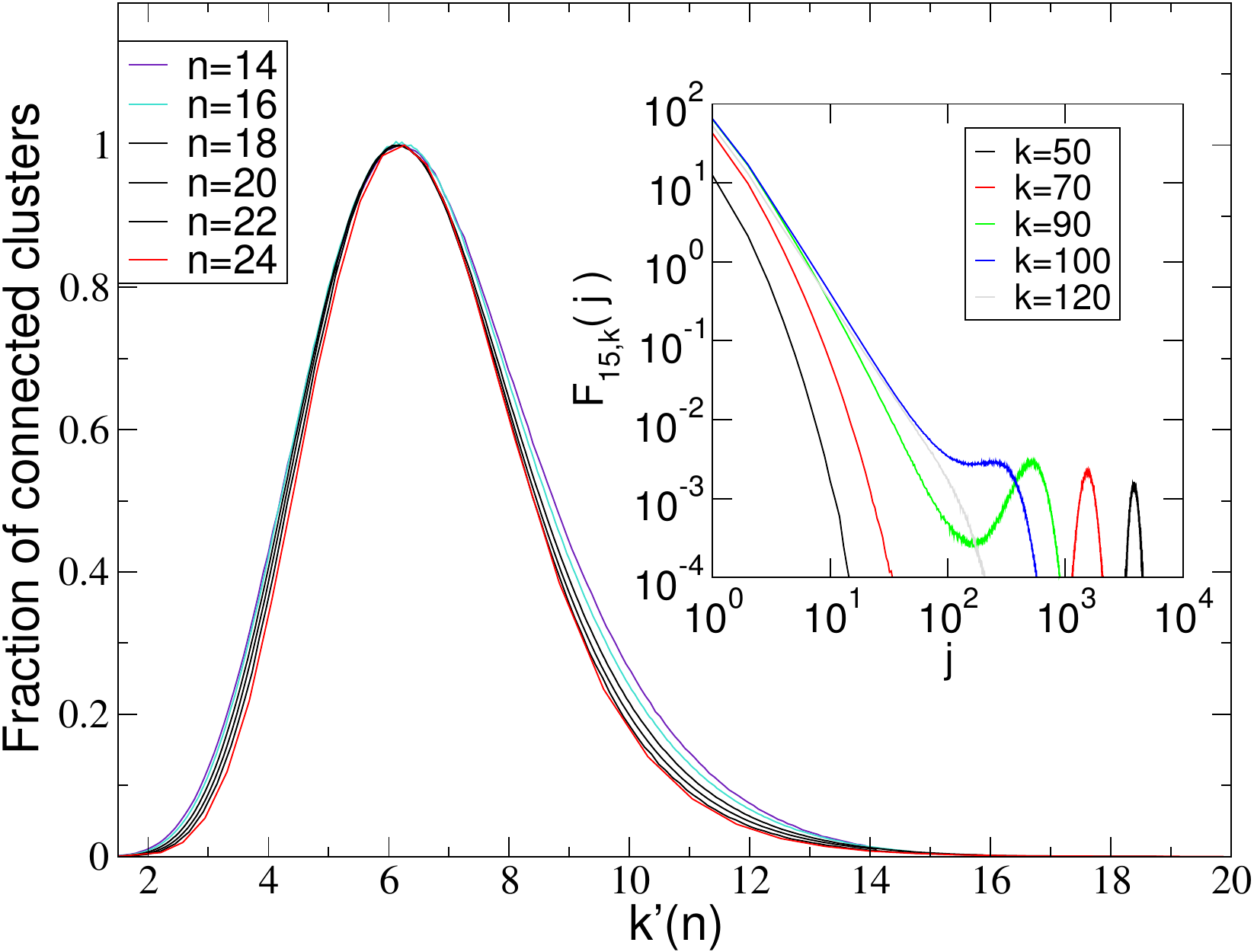}
\includegraphics[width=0.86\columnwidth]{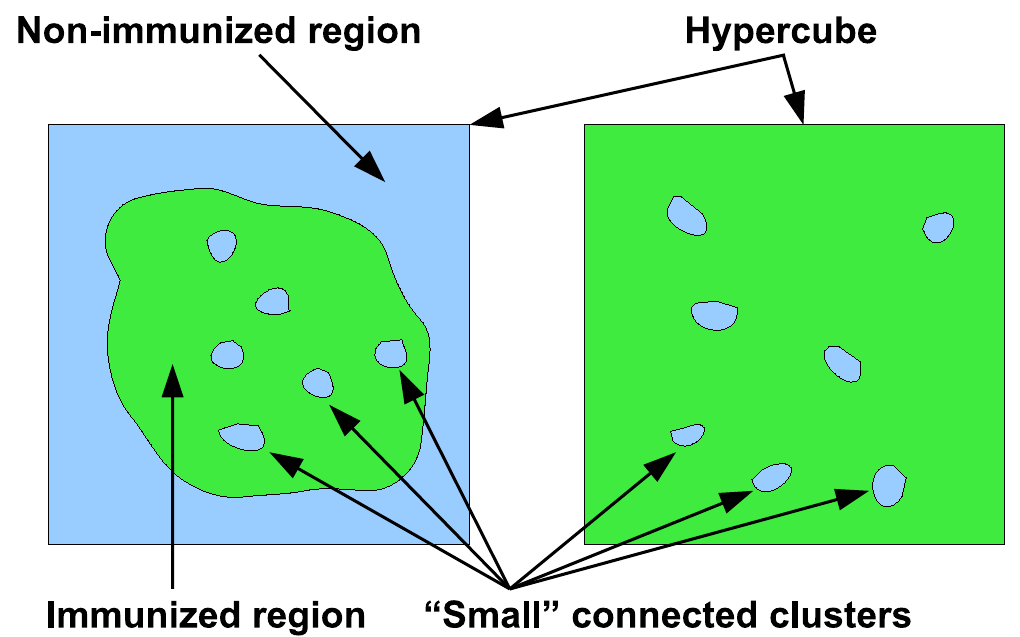}
\caption{ {\bf Top} Average number of connected clusters,
  $F_{15,k}(j)$, composing the CEIS as a function of $k^{\prime}(n)$
  as defined in the text. For each value of $n$ the functions plotted
  have been divided by their maximum value. The functions collapse
  into a well defined function, independent of $n$, as $n$ increases.
  {\bf Inset.} Distributions $F_{15,k}(j)$ for several values of $k$.
  For $k<90$ the functions feature two disjoint peaks with the
  rightmost peak, whose area is equal to $1$, moves to the left as $k$
  increases. For $k>90$ the two peaks merge. {\bf Bottom} Sketch of
  the topology of the immunized (green) and non-immunized (blue) region
  of the sequence space. \textit{Left}: the set CEIS is composed by
  one {\em big} connected cluster, corresponding to a non-immunized
  region of the hypercube, and many {\em small} connected clusters,
  corresponding to ``holes'' contained into the EIS. \textit{Right}:
  the whole hypercube is immunized apart from a few non-immunized
  clusters.}
\label{fig:collassi}
\end{figure}

In the main text we have studied the cluster structure of the EIS. Let
us first recall that the immunity set, $I_n(\vec{v})$, of a single
string $\vec{v}$ is a connected set. Without loss of generality we
consider $\vec{v}=\vec{0}$ since the $I_n(\vec{v})$ is invariant under
translation on the hypercube with periodic boundary conditions,
through the translation operator (\ref{math:translationoperator}). We
can think $I_n(\vec{0})$ as the union of the disjoint sets,
$I_n(\vec{0})= \cup_{i=0}^n N_i$, with

\begin{equation}
N_i \equiv \{\ z \in I_n(\vec{0}) : d_h(\vec{0},\vec{z})=i \ \}.
\end{equation}

\noindent For each $\vec{z}_i \in N_i$ there is always a nearest
neighbor contained in $N_{i-1}$ such that
$d_h(\vec{z}_i,\vec{z}_{i-1})=1$. This implies that, for each string
in $I_n(\vec{0})$, it always exists a sequence of nearest neighbors to
connect that string to $\vec{0}$, i.e. $I_n(\vec{v})$ is a connected
set.

\noindent Starting from this result, in the main text we have shown
that the EIS is always connected, though not simply connected. In
fact, when $k$ strings are drawn at random, there exists a threshold
for $k=\lceil n/2\rceil $ above which the complementary EIS (CEIS) can
be broken down in clusters. This is due to the fact that we need to
choose at least $\lceil n/2\rceil$ strings in order to generate an EIS
that contains ``holes'', as it is shown in the sketch in Figure
\ref{fig:collassi} (bottom). An example is given by the following
infection set:
$$
\begin{array}{c c c c c c}
  (\, 1,\, 1,\, 0,\, 0,\, 0,\, 0 ) \\ 
  (\, 0,\, 0,\, 1,\, 1,\, 0,\, 0 ) \\ 
  (\, 0,\, 0,\, 0,\, 0,\, 1,\, 1 ) \\
\end{array} 
$$

\noindent The string $(0, 0, 0, 0, 0, 0)$ is not contained into the
EIS generated by this set, on the contrary of all its neighbours.
Therefore, the string alone constitute a cluster of the CEIS. However,
other infection sets can generate CEIS featuring a much more complex
cluster structure.

Figure~\ref{fig:collassi} (top) shows the average number of connected
clusters in the CEIS (divided by the maximum value for each $n$) as a
function of the rescaled variable:

\begin{equation} {k}^{\prime}(n)=\frac{k}{2^{ \eta n}} \cdot
  \frac{1}{1 - 2^{ - \gamma \cdot n}},
\end{equation} 

\noindent where the exponent $\gamma \simeq 0.12$ estimates the finite
size scale effects and $2^{-\eta n}$ is the fraction of strings
contained into the immunity set of a single strain with
$\eta=1-\ln_2{\phi} \sim 0.306$. Therefore the normalized number of
connected clusters in the CEIS can be rescaled on a single master
curve as $n$ increases, thanks to a suitable rescaling of $k$.

A more detailed description of the cluster structure of the CEIS can
be obtained by investigating the dependence on $n$ and $k$ of the
average size of the connected components composing it. To this end we
define the distribution functions $F_{n,k}(j)$ as follows.
$F_{n,k}(j)$ are defined as the average number of connected clusters,
with cardinality $j$, generated by an infection set with $k$ 
randomly drawn strains. In Fig~\ref{fig:collassi} (inset of the
top panel) we report the distributions for $n=15$ and several values
of $k$. It turns out that, for values of $k$ not too large, the
complementary set is composed by one big connected cluster and many
small connected clusters. In fact the distributions exhibit two
disjoint peaks: one centered on a large value of $j$, due to the
contribution of the {\em big} cluster, and one centered on small
values of $j$, given by the contribution of the {\em small} clusters.
Further analysis reveals that the area of the former peak is always
equal to $1$ and that for every choice of the infection set there is
always only one big cluster. On the other hand, the number of the
small clusters is not fixed and depends on the infection set. On the
contrary, for larger values of $k$, only the small connected clusters
remain: increasing $k$ the rightmost peak disappears, moving on the
left and merging with the leftmost peak. As sketched in
Fig.~\ref{fig:collassi} (bottom), it is reasonable to interpret the
big cluster as a region of the hypercube which has not been immunized
yet and the small connected clusters of the CEIS as {\em holes}
contained into the immunized region of the hypercube. This cluster
structure of the CEIS could have a strong impact on the underlying
virus-host interaction, which could be investigated through a more
realistic simulation of the virus-host dynamics.

\end{document}